\documentclass[]{llncs}

\usepackage[utf8]{inputenc}
\usepackage{amssymb,amsmath}
\usepackage{graphicx,tabularx}
\usepackage[labelfont=bf,size=small,labelsep=period]{caption}
\usepackage[labelfont=default]{subcaption}
\usepackage{color}
\usepackage{mathptmx}
\usepackage{wrapfig}
\usepackage{paralist}
\usepackage{enumitem}
\usepackage{hyperref}
\usepackage[linesnumbered,algoruled,longend,vlined]{algorithm2e}
\usepackage{xspace}
\usepackage{multirow}
\usepackage[figuresright]{rotating}
\usepackage{todonotes}
\usepackage{microtype}

\newcommand{\pcenter}{\ensuremath{p_\mathrm{c}}\xspace}
\newcommand{\lunit}{\ensuremath{l_\mathrm{unit}}\xspace}
\newcommand{\frep}{\ensuremath{F_\mathrm{r}}\xspace}
\newcommand{\fatt}{\ensuremath{F_\mathrm{a}}\xspace}
\newcommand{\fedge}{\ensuremath{F_\mathrm{e}}\xspace}
\newcommand{\fgrav}{\ensuremath{F_\mathrm{g}}\xspace}
\newcommand{\ladj}{\ensuremath{l_\mathrm{adj}}\xspace}
\newcommand{\lnadj}{\ensuremath{l_\mathrm{nadj}}\xspace}
\newcommand{\cdeg}{\ensuremath{c_{\deg}}\xspace}
\newcommand{\pframe}{\ensuremath{p_\mathrm{f}}\xspace}
\newcommand{\fframe}{\ensuremath{F_\mathrm{f}}\xspace}
\newcommand{\wmin}{\ensuremath{w^\star}\xspace}
\newcommand{\hmin}{\ensuremath{h_{\min}}\xspace}
\newcommand{\wimin}{\ensuremath{w_{\min}}\xspace}
\newcommand{\clen}{\ensuremath{c_\mathrm{len}}\xspace}
\newcommand{\cpre}{\ensuremath{c_\mathrm{pre}}\xspace}
\newcommand{\frepbezier}{\ensuremath{F_\mathrm{rb}}\xspace}
\newcommand{\fattbezier}{\ensuremath{F_\mathrm{ab}}\xspace}
\newcommand{\nmax}{\ensuremath{n_\mathrm{max}}\xspace}

\graphicspath{{images/}}

\begin{document}
\date{}

\title{Drawing Graphs within Restricted Area}

\author{Maximilian Aulbach \inst1 \and Martin Fink \inst2
  \thanks{The work of Martin Fink was partially supported by a
fellowship within the Postdoc-Program of the German Academic Exchange Service
(DAAD).} \and Julian
Schuhmann \inst1 \and Alexander Wolff \inst1}

\institute{Lehrstuhl f\"{u}r Informatik I, Universit\"{a}t
W\"{u}rzburg, Germany \and Department of Computer Science,
University of California, Santa Barbara, USA}

\maketitle
\begin{abstract}
  We study the problem of selecting a maximum-weight subgraph of a
  given graph such that the subgraph can be drawn within a prescribed
  drawing area subject to given non-uniform vertex sizes. We develop
  and analyze heuristics both for the general (undirected) case and
  for the use case of (directed) \emph{calculation graphs} which are
  used to analyze the typical mistakes that high school students make
  when transforming mathematical expressions in the process of
  calculating, for example, sums of fractions.
\end{abstract}

\section{Introduction}
\begin{figure}[h]
	\centering
	\includegraphics[scale=.8]{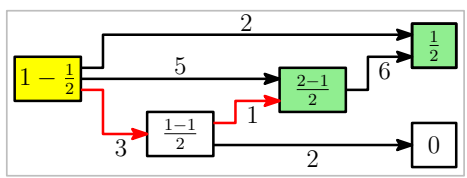}
	\caption{Example of a calculation graph. Edge weights are indicated
	on the edges.}
	\label{fig:calc-graph-example}
\end{figure}

Our motivation for the problem that we study in this paper stems from
so-called \emph{calculation graphs}.  Calculation graphs
represent calculations starting from some initial task.  They
are used in studies~\cite{hennecke07} involving large numbers of high school
students in order to analyze the students' typical mistakes in elementary
mathematics. Even for relatively simple tasks such as evaluating the
term ``$3 \cdot (2 + 1/5)$'', the different transformations
performed by a large number of subjects can result in calculation
graphs with hundreds of vertices. With the help of drawings of
calculation graphs, human experts should be able to analyze how
students calculate and, especially, which mistakes they frequently
make. As Hennecke~\cite{hennecke07} suggests, such drawings
are only useful if they are not too large, that is, if
they fit into a relatively small drawing area. Hence, well-readable
drawings of important parts of the graphs must be generated in an
automated fashion. 
Certainly, the drawn subgraph should represent as much information
of the original calculation graph as possible. 
Therefore, we consider the graphs to be edge- and
vertex-weighted; see Fig.~\ref{fig:calc-graph-example} for an
example.
The weight of an edge is the number of students who
applied the respective calculation step; the weight of a vertex is the
number of students who had the given term as an intermediate result in
their calculation.
Since we want the labels of the vertices to be readable, we assume
that their sizes are fixed.  Hence, often only a small fraction of the
graph will fit into the prescribed drawing area.
Note that a user of such a drawing must be made aware that
only a subgraph is shown.

Certainly, vertices and edges that occur
only in a single student's calculation can easily be dropped. For
higher weights, however, this is not as easy. For instance, it could
be possible that dropping a single vertex of medium weight allows us
to include several other vertices in the drawing, each having almost
the weight of the dropped vertex; the resulting drawing can therefore
allow a better analysis. Hence, we develop a method for
selecting the drawn subgraph based on the graph structure and not only
on the weights.

\paragraph{Further Examples.} Apart from calculation graphs, there are
also other scenarios in which most of the information of a large graph
should be presented in a limited area to the user. A first example are
social networks, which are usually very large. In this setting, both
labels of the vertices and weight (e.g., given by the person's
activity in the network) occur. By drawing a heavy subgraph in a
restricted area, a user can quickly get an overview over important
actors in the network and the connections between them. Another
example, that we will use in the main part of the paper, are coauthor
networks, where the weights represent numbers of publications. Here,
we want to find a drawing of a subgraph that represents as many of the
publications as possible, i.e., we prefer to keep authors with many
publications.

\paragraph{Related Work.} 
Surprisingly, the very natural problem of selecting a subgraph that
can be drawn within a prescribed area seems to be new.  Some work on
related problems, however, is worth being mentioned.  Rather than
fixing the drawing area and maximizing the size of the subgraph that
fits into this area, graph drawing research has focussed on drawing
the whole input graph and minimizing the area needed for the
drawing, which has also been the task in graph drawing
contests~\cite{gdcontest2012}. This
problem is known to be NP-hard for orthogonal drawings with given
orthogonal representation (that is, fixed
bends)~\cite{Patrignani200147}.  In constrained graph
layout~\cite{He1998,Dwyer2009}, the user can constrain the region into
which a vertex must be placed.  
Dwyer et al.~\cite{ipsepcola} presented the algorithm IPSep-CoLa that
allows constraints demanding a vertical or horizontal separation of
vertex labels. Especially, the constraints allowed in their algorithm
make it possible
to enforce that vertices are placed within a prescribed area. However,
in contrast to our work, they do not consider dropping vertices. If
the vertices do not fit into the area, their algorithm will not find a
feasible drawing.
The well-known algorithm of Fruchterman and
Reingold~\cite{Fruchterman1991} for force-directed graph drawing
allows to specify a rectangular area within which the whole graph has
to be drawn. However, in contrast to our work, the algorithm can
easily achieve this since vertices are drawn as points without label
and edges can be made arbitrarily short, which we do not allow.

Dwyer et al.~\cite{dwyerea2008overview+detail} developed a method for
interactive exploration of graphs based on constrained graph layout.
In their method, they use a fast heuristic for the overview drawing;
for the detailed view of a smaller part of the graph, however, they
can afford to use a slower constrained graph layout algorithm that
yields better results. Constraints are used in order to ensure
consistency between the views and preserve the users's mental map.
Da Lozzo et al.~\cite{JGAA-252} considered the problem of drawing
graphs on the very small display of a smartphone. In contrast to the
approach used in this paper, the did not try to represent the whole
graph on the display, but rather decided to provide only a local view
around a focus vertex; their approach then offers interactive ways of
navigating through the graph and exploring the graph based on the
local view.

%

\paragraph{Our contribution.} 
While calculation graphs have quite specific characteristics and
drawing requirements (see Section~\ref{sec:calculation}), the problem
of graph drawing under area restrictions is also of interest for
general, undirected graphs.  For both cases, we present heuristics
and evaluate them experimentally.

For the general case, we use the force-directed approach, reusing
forces defined by Fruchterman and Reingold~\cite{Fruchterman1991} and
by Bertault~\cite{Bertault2000}.
We add extra phases in which the graph is pressed together;
from time to time we remove vertices or edges from the graph if this
is needed for further compaction. We use the desired edge length as
parameter both for the usual iterations and our extra phases.
In our tests, this proved to be effective for parametrizing
the density of the output drawing. Furthermore, we experimentally
improved the way of avoiding vertex-edge intersections and we
developed a postprocessing that routes some of the edges as curves.

For calculation graphs, we chose the well-known Sugiyama framework
\cite{sugiyama1981methods} as basis
for our algorithm since we want calculation paths to be readable
from left to right. We add a method that successively removes the
least important vertices and edges until the drawing fits into the
given area.  We also consider the
weight of edges for crossing minimization so that important edges have
few crossings. In our tests, it turned out that removing the lightest
vertices as a preprocessing often improved the weight of the final
subgraph. Furthermore, routing the edges as curves gave very nice
results, also compared to the original orthogonal drawing style for
calculation graphs.

\section{General Graphs}
\label{sec:general}

We first present an algorithm for drawing arbitrary graphs.
The input of our problem consists of an unweighted graph $G =
(V,E)$ with a weight function $w \colon V \cup E \to
\mathbb{R}^+$. For each vertex $v \in V$, we are given a geometric
object $\ell(v)$ that will represent the vertex in the drawing.
Vertices can be represented by different shapes, e.g., rectangles,
disks, or ellipses. We will focus on rectangular vertices which are
well-suited for text labels of vertices. We will denote the height
and the width of $\ell(v)$ by $h_v$ and $w_v$, respectively.

In addition to the graph input, we are given an axis-parallel
rectangle of height~$H$ and width~$W$, the \emph{drawing area}. The task is
to find a subgraph $G' = (V',E')$ of $G$ with a \emph{nice} drawing of
$G'$ within the given drawing area.

The hard constraints for the drawing are clear: Each vertex $v\in V'$
must be represented by $\ell(v)$, the vertices must not overlap, and
each edge must connect its incident vertices. However, it needs to be
clarified, what a nice drawing is. In our setting with a
restricted drawing area, putting vertices close together can
allow us to have more vertices---and thus more weight---in the final
drawing. Certainly, a drawing with vertices that are very close
is not nice; the same holds for very short edges. We will
discuss these criteria and more later in detail.

By a straightforward reduction from \textsc{Subset Sum}, where we use
height 1 for all vertices and the drawing area, we can easily
observe that maximizing the weight of the subgraph that can be drawn
in a prescribed area is NP-hard.

\subsection{Our Algorithm}
\label{sec:gen-algo}
Our approach makes heavy use of the force-directed framework; in this
class of graph drawing algorithms, the drawing is incrementally
improved, starting with an arbitrary layout. Each improvement step is
done by letting \emph{forces}, defined using physical analogies, move
the vertices.

In contrast to usual force-directed algorithms, we have to take both
the dimensions of vertices and of the prescribed drawing area into
account. Therefore,
we add two important ingredients: We try to fit the drawing into a
frame of decreasing size, and we remove vertices or edges from
the graph in order to make the graph smaller so that the current
drawing can fit into the current frame. While, as a first idea,
fitting the drawing into the given area could be steered by a force
pulling all vertices towards the center of the drawing region, this
idea has some drawbacks. Therefore, we will introduce a more advanced
approach. Similarly, removing vertices could simply be done by removing
the lightest vertex in each step. However, this would take neither
the structure of the graph nor the current drawing into account.
Hence, we introduce a measure for the \emph{stress} of vertices in the
current situation; we will always remove the vertex with the highest
stress value. Algorithm~\ref{alg:general-graphs} in
Appendix~\ref{app:gen-tables} outlines the basic structure of our
algorithm. In the following paragraphs, we will detail out the
individual steps.

\paragraph{Forces.}
\label{sec:gen-forces}
We first define the forces used in our algorithm.
\begin{compactitem}
  \item We reuse existing forces from the algorithm of Fruchterman and
    Reingold~\cite{Fruchterman1991}; that is, for any pair $u,v \in V$ of
    vertices, there is a force $\frep\left(u,v\right)=
    \lunit^{2}/(d\left(u,v\right))\cdot\overrightarrow{uv}$ on~$v$
		that repels~$v$ from~$u$, where $\overrightarrow{uv}$ is the unit
		vector pointing from~$u$ towards~$v$. If the vertices are
    adjacent, that is, if $uv \in E$, there is an additional force
		$\fatt(u,v)= (d(u,v))^{2}/\lunit\cdot\overrightarrow{v
		u}$ that attracts $v$ towards $u$.
		Both forces use a factor \lunit which describes the
		desired unit edge length. Since the desired edge length heavily
		influences the density of the drawing, and, hence, the number
		and weight of vertices and edges that can be placed within the
		given drawing area, the choice of $l_{\mathrm{unit}}$ is crucial
		for the results. While we allow the parameter to be set freely,
		we stress that the value must be chosen carefully, taking the
		sizes of vertices into account, so that one gets nice output
		drawings.
  \item Due to the high density of the input graphs and the given
    sizes of vertices, it may easily occur that vertices are
    intersected by nonincident edges, which reduces the readability
    significantly. As a first step to overcome this problem, we use a
    force that has been introduced by Bertault in his PrEd
    algorithm~\cite{Bertault2000}: If an edge~$\left\{ u,w \right\}$
    intersects a vertex~$v$ in its inner region, that is, close to the
    center of $v$, then we let a
    force~$\fedge\left(v,\left\{u,w\right\}\right)=
    {\left(\lunit-d\left(v,i_v\right)\right)}^{2} \cdot
    \overrightarrow{i_vv}$ repel $v$ from $\left\{ u,w
    \right\}$, where the point $i_v$ is the orthogonal projection of
		$v$ onto the straight-line segment $\overline{uw}$. Note that this does not guarantee
		that intersections between vertices and edges are avoided.
		However, such intersections become less likely; in
		Sec.~\ref{sec:gen-ext}, we will see how their number
		can be further reduced by routing edges as curves.
  \item For making the drawing more compact, we introduce a force
    $\fgrav(v)= d\left(v,\pcenter\right) \cdot
    \overrightarrow{v\pcenter}$ that attracts each
		vertex~$v$ to the center~\pcenter of the drawing area.
		In our experiments it turned out that activating this force in
		later steps of the algorithm reduces the time for finding a final
		drawing, but also the quality of the output. Therefore, as a
		default, the force is only active when computing the very first
		equilibrium layout.
\end{compactitem}

\paragraph{Handling the Frame.}
\label{sec:gen-frame-handling}
The forces described above yield a functional
force-directed algorithm which can be applied for getting an
initial layout. Once we have an initial drawing, we initialize the
frame~$F$ as the bounding box of the drawing. Our algorithm
iteratively reduces the size of the frame until it matches the
prescribed drawing area.

In each step, we first uniformly reduce the height and the width
of~$F$ by a small amount. It may happen that some vertices (partially)
lie outside of the resulting new frame~$F'$. If this is the case for a
vertex~$v$, we just place it at the closest position that lies
completely within~$F'$. This operation can result in intersections of
vertices. Therefore, we compute a new equilibrium state which
hopefully solves the intersections.

In all force-directed iterations in which there is a frame, we
will always ensure that no vertex leaves the frame. This is done by
cutting off the resulting movement vectors; Fruchterman and
Reingold~\cite{Fruchterman1991} did the same for ensuring that no
vertex leaves the drawing area in their algorithm---with the
difference that they did not shrink the frame but rather started with
a very compressed drawing since in their setting vertices are points
that can be arbitrarily close.

If there are intersections after computing a new equilibrium layout,
we have a clear indication that the graph is still too large for the
current frame and, hence, for the desired drawing area. In this case,
and also in some more cases, we will remove vertices or edges as
described in the next section.

\paragraph{Removing Vertices and Edges.}
\label{sec:gen-remove}
Our indicator for the necessity of the removal of vertices or edges
is, roughly speaking, the density of the drawing. If there are too
many vertices in the graph for the current frame, then vertices will
come very close. Therefore, we decide to remove a vertex or an edge if
the minimum edge length is less than a value \ladj or if the minimum
distance between two nonadjacent vertices is less than a value \lnadj.
Note that this includes the case of two intersecting vertices.

Deciding what should be removed is more difficult. Since we
want to keep as much weight as possible, the natural idea is to
remove the lightest vertex. However, this is often not the best
choice---even if it is unambiguous. A more advanced
criterion should take also the degree of a vertex into account. The
higher the number of neighbors of a vertex is, the more information on
the graph is lost by removing the vertex.

\paragraph{Stress Calculation.}
However, we can still do better: So far, we have not taken the current
drawing into account, although it can give us valuable information
about the density of vertices in the vicinity of the vertex that
should be removed. Thus, we suggest doing so by considering the forces
in the last equilibrium layout. Even if the total movement vector of a
vertex has length zero, this may actually result from strong forces
that try to move the vertex to different directions, e.g., if the
vertex is ``trapped'' between many other vertices around it.

\begin{wrapfigure}[12]{r}{.32\textwidth}
	\centering
	\includegraphics[scale=.87]{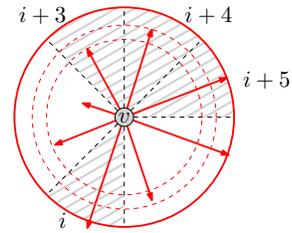}
	\caption{Pressure computation for one of the octants.}
	\label{fig:pressure-octants}
\end{wrapfigure}

These considerations lead us to a measure that we call the
\emph{pressure} of a vertex. Intuitively, the pressure is the maximum
strength of forces in roughly opposite directions that act on a
vertex. For formalizing this, we subdivide all force vectors applied
on the vertex~$v$ into eight octants. For each
octant, we sum up the force vector. For $i=0, \ldots, 7$ let
$l_i$ be the length of the resulting force vector for octant~$i$.
Now, we first build the pressure~$p_i$ for octant $i$ by comparing $l_i$ with
the force vectors in the three opposite directions, i.e., with
$l_{i+3}$, $l_{i+4}$, and $l_{i+5}$ ($\mathrm{mod}~8$); see
Fig.~\ref{fig:pressure-octants}.
The pressure then is
the maximum over the pairwise minima, i.e., $p_i = \max \left\{
  \min\left\{ p_i, p_{i+3} \right\}, \min\left\{ p_i, p_{i+4}
\right\}, \min\left\{ p_i, p_{i+5} \right\} \right\}$. The total
pressure on~$v$ is $p(v) = \max \left\{ p_0, \ldots, p_7 \right\}$.

Now, we must integrate the weight $w(v)$ and the degree
$\deg(v)$---indicators of the vertex's resistance against
pressure---with the pressure in order to get the \emph{stress}~$s(v)$.
We do so by setting $s(v) = p(v)/(w(v) \cdot (\deg(v) +
\cdeg ))$. Here, \cdeg is a small positive constant that ensures that
we do not get problems for isolated vertices and that steers our preference
for keeping isolated vertices in the drawing. With this definition of the stress, we
can always choose the vertex with the highest stress for removal.

\paragraph{Boundary Vertices.} There is a special case for vertices
close to the boundary of the frame. We never move a vertex over
the boundary although, in many cases, vertices are repelled in this
direction by the inner vertices. This phenomenon is a cause of
pressure on a vertex that is, so far, not covered by our definition.
We therefore introduce a new ``virtual'' force that mimics the
resistance against movements that push the vertex out of the frame.
This force repels the vertex perpendicularly to the inside of the
frame, away from the closest point $\pframe$ of the frame's boundary. More
precisely, the virtual force is~$\fframe(v) =
\overrightarrow{\pframe v} \cdot \lunit^2/d(v,\pframe)$.
We stress that $\fframe$ is only taken into account for the stress
computation and not actually applied.

\paragraph{Edges.}
In some cases, one may try to remove only an edge instead of a whole
vertex; the hope is that after removing the edge, the graph
becomes more flexible so that the available space can be used better.
As an indicator for such a situation, we use the average edge length
in the current drawing. If this length is larger than $\lunit \cdot
\clen$, with a factor $\clen > 0$, we decide for removing an edge
instead of a vertex. The intuition is that, on average, the edges are
not very short, which means that by removing one of these longer edges
we could allow more flexibility to the placement of vertices.

In order to determine which edge will be removed, we, again, use a
definition of stress. To this end, we take both the
weight~$w(e)$ and the weights of the edges crossing~$e$ into account.
Let $E'$ be the set of these edges. Then, we set $s(e) =
\sum_{e' \in E'} w(e') \cdot |E'|/w(e)$ to be the stress
of~$e$, and we remove the edge with the highest stress value.

\subsection{Extensions}
\label{sec:gen-ext}
We developed and implemented two extensions that can help improve the
runtime of the algorithm and the quality of the resulting drawings,
respectively.

\paragraph{Preprocessing.}
In many input instances, there is a large number of vertices with
very small weight, for which it is very unlikely to occur in the final
drawing. To speed up the algorithm, we can remove all vertices that
are lighter than a threshold value \wmin. Our choice of $\wmin$ is
based on guessing a bound on the maximum number of vertices in the
final drawing and depends on the height~$H$ and the weight~$W$ of the
drawing area, the minimum height~\hmin and the minimum width~\wimin of
a vertex as well as on the desired edge length~\lunit
and a factor $\cpre > 0$. We will make sure that we keep at least
$(H \cdot W)/\left( (\lunit \cpre + \hmin) \cdot (\lunit \cpre +
\wimin) \right)$ vertices in the graph.

\paragraph{Postprocessing: B\'ezier Curves.}
In Sec.~\ref{sec:gen-algo}, we explained the force that
aims at avoiding intersections between vertices and nonincident edges.
However, we cannot guarantee that we do not have such intersections.
For the final drawing, we can do two things about this: (i)
Intersections of an edge with the outer region of a vertex are
relatively easy to distinguish from incidences and can, therefore, be
tolerated. (ii) We can remove more edges if necessary.

Here, we present a third possibility: If we allow edges to be curves
rather than only straight-line segments, we can avoid more
intersections.
In their ImPrEd algorithm~\cite{Simonetto2011}, an improvement to the
work of Bertault~\cite{Bertault2000}, Simonetto et~al.\ allowed edges
to be drawn as polylines of different complexity, where bends were
introduced and removed during the algorithm, based on the current
drawing. However, this approach made the algorithm much slower; also,
polylines with sharp bends are not easy to follow. Therefore, we
use a postprocessing in which edges are routed as B\'ezier curves
around intersected nonadjacent vertices.
We do this by representing edge $e = \left\{ u,w \right\}$ by a
quadratic B\'ezier curve, i.e., a parametric curve with a control
point~$p_e$ in addition to the endpoints.

\begin{figure}[tb]
	\begin{subfigure}[t]{.38\textwidth}
		\centering
		\includegraphics[page=1]{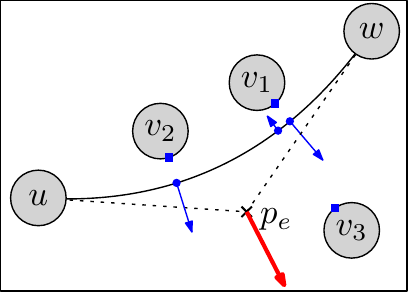}
		\caption{Repelling forces}
		\label{fig:gen-bezier-rep}
	\end{subfigure}
	\hfill
	\begin{subfigure}[t]{.58\textwidth}
		\centering
		\includegraphics[page=2]{gen-bezier-rep}
		\caption{Attracting forces; $v_3$ is too far away for an
		attraction.}
		\label{fig:gen-bezier-attr}
	\end{subfigure}
	\caption{Repelling and attracting forces for B\'ezier curves.}
	\label{fig:gen}
\end{figure}

The computation of the curve is done as a postprocessing in the very
last step of the algorithm. It is realized as an additional
force-directed algorithm in which only the control point is moved,
starting at the position in the middle between $u$ and $w$. Each
vertex~$v$ that is not far away from $e$ causes a repelling force on
$p_e$; see Fig.~\ref{fig:gen-bezier-rep}. This force is parametrized by
the width $w_v$ and the height $h_v$ of $v$ as well as by the point
$p_v'$ of $v$ that is closest to $e$ and the point $p_e'$ of
$e$ that is closes to $v$. The repelling force is defined as
$\frepbezier(e,v) = \overrightarrow{p_v' p_e'} \cdot (w_v^2 +
h_v^2)/d(p_v', p_e')$.

In order to avoid that the edge is curved too
much, we also have an attracting force for vertices that have been
(almost) intersected by $e$. If $v$ is such a vertex, then the
attracting force $\fattbezier(e,v) = \overrightarrow{p_e' p_v'} \cdot
d(p_v', p_e')^2/\sqrt{w_v^2 + h_v^2}$ is applied to~$p_e$; see
Fig.~\ref{fig:gen-bezier-attr}.

\subsection{Experiments and Evaluation}
We implemented our heuristic in Java, using the graph library
JUNG\footnote{http://www.jung.sourceforge.net/}. Our experiments were
performed on a Core i5-2500K CPU with 8 GB of RAM. For the
force-directed part of the algorithm, we also used a cooling factor
that slows down the movement of vertices over the iterations in order
to accelerate the computation of an equilibrium. We configured our
algorithm such that there are always 25 steps of shrinking the frame
around the current drawing. As input data,
we primarily used (subgraphs of) the graph drawing collaboration graph
from 1994 till 2012; in total, the graph has 950 vertices and 2559
edges. The weight of each vertex is the number of
publications of the respective author and the weight of an edge
connecting two authors is the number of their joint publications.
We focused on a drawing area of $29.7 cm \times 21 cm$ (DIN A4) where
the size of vertices was determined by the author's name (in 10pt
font). For the complete publication graph, the runtimes were about 2
minutes, depending on the parameters.

The main results of our tests are the following.
\begin{compactitem}
  \item The preprocessing step described in Sec.~\ref{sec:gen-ext}
    pays off; depending on the parameter \cpre, we may save
    runtime and get better results. In our tests, $\cpre = 0.7$ seemed
		a good choice; see Table~\ref{tab:cpre} in
		Appendix~\ref{app:gen-tables}.

	\item For testing whether something has to be removed, we set
    $\ladj = 0.1 \lunit$ and $\lnadj = 0.15 \lunit$. For the parameter
    $\clen$ that determines whether an edge or a vertex will be
    removed, $\clen = 0.9$ gave a good compromise between
    the vertex and the edge weight in the final drawing; see also
		Table~\ref{tab:clen} in Appendix~\ref{app:gen-tables}.
  \item We tested the effect of not always activating the force that repels
    vertices from edges; instead, we first computed an
    equilibrium without the force and then another one with it, so
    that we have more flexibility for vertices to cross edges. In our
    tests, this proved to have a significant impact on the total
    weight of edges in the final drawing, yielding an increase of
    almost 80\%; see Table~\ref{tab:doubleCalculation} in
		Appendix~\ref{app:gen-tables}.

  \item The following way of computing the total force
    vector $F$ for a vertex based on putting weights to the single
    forces showed the best results: $F = 0.01 \frep + 0.01 \fatt +
    0.005 \fgrav + 0.0075 \fedge + 0.01 \fframe$. Note that not
    all forces are active at the same time.
\end{compactitem}

\begin{figure}[tb]
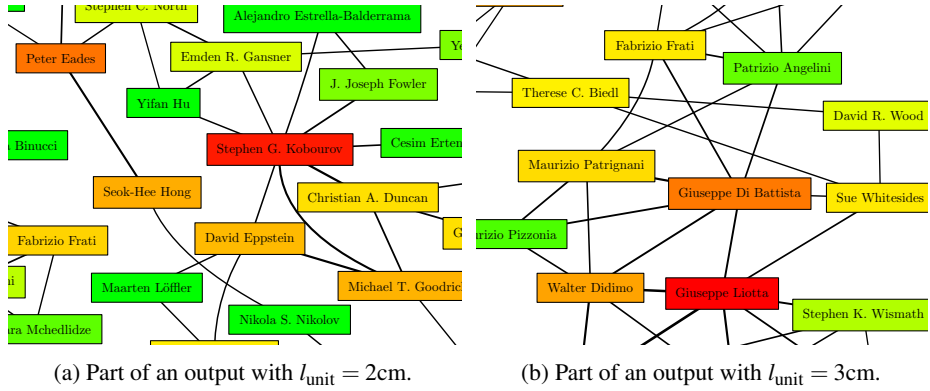

  \begin{subfigure}[t]{.49\textwidth}
    \centering
    \includegraphics[scale=.5,clip=true,trim=9.7cm 4cm 8cm 8cm]{result1}
    \caption{Part of an output with $\lunit = 2$cm.}
    \label{fig:gen-result1-crop}
  \end{subfigure}
  \hfill
  \begin{subfigure}[t]{.49\textwidth}
    \centering
    \includegraphics[scale=.5,clip=true,trim=17.55cm 4cm 0.2cm 8cm]{result2}
    \caption{Part of an output with $\lunit = 3$cm.}
    \label{fig:gen-result2-crop}
  \end{subfigure}
  \caption{Output examples with different choices for \lunit.}
  \label{fig:gen-results-examples-crop}
\end{figure}

A central parameter used in our algorithm is \lunit, describing the
desired edge length. Figure~\ref{fig:gen-results-examples-crop} shows
output examples that demonstrate that higher values lead to
drawings that are less dense. Full outputs are attached in
Appendix~\ref{app:gen-output}.

\section{Calculation Graphs}
\label{sec:calculation}

We now consider the initial problem of drawing calculation graphs in
prescribed area.
In our terminology, most of the input stays the same compared to
general graphs, except that now our edges are directed since they
represent calculation steps. Additionally, we are given a
\emph{start vertex} $s \in V$ that represents the task given to the
students. Hence, $s$ must be present in the output drawing and,
furthermore, we insist that in the subgraph that is finally drawn,
each vertex can be reached from $s$. To further improve the
readability of the steps, we want as many edges as possible pointing
from left to right. As a drawing convention for the edges, we will use
the orthogonal drawing style with edges leaving vertices horizontally;
we try to minimize the numbers of crossings and of bends and to
optimize the gaps between edges and vertices.
The hardness proof for general graphs can easily be adjusted. Hence,
the problem remains NP-hard.

\subsection{Our Algorithm}
Due to the required readability from left to right, our approach is
based on the Sugiyama framework~\cite{sugiyama1981methods} for
hierarchical graph drawing. This framework consists of several steps
that we will adjust to our problem.  Our adjustments are optimized for
our drawing style and especially for the task of removing vertices or
edges from the graph, if necessary. We first briefly review the steps
of the Sugiyama framework before going into detail and describing our
adjustments.

In the first step, the graph is made acyclic by reverting some of the
edges. Next, the vertices are assigned to layers from left to right.
Then, based on the layer assignment, the number of edge crossings is
minimized; this step results in relative
orders of the vertices of each layer. Eventually, the final
coordinates of vertices are computed and the edges are routed.

\paragraph{Breaking the Cycles.}
In the later steps of the Sugiyama framework, it is assumed
that all edges are directed from left to right, i.e., the graph must
be acyclic. To ensure this, we
have to revert some edges. If we want to  revert the smallest number
of edges, we have to solve the NP-hard Feedback Set problem. We can either
use existing heuristics, or even afford finding an
optimum solution with the help of an MIP solver, which in our tests
worked quite fast; this allows us also to minimize the weight of the
reverted edges rather than their number.

\paragraph{Layer Assignment.}
Several approaches for the layer assignment in the Sugiyama framework
exist, depending on the objective, e.g., of minimizing the number of
layers or the number of vertices in a layer. Often,
the number of layers is minimized subject to a prescribed maximum
height of each layer. However, we will not
use the height of our drawing area as the maximum height of a layer,
although, at first, this may seem a good idea: If we do so, we would
most probably have to remove many layers of vertices completely from
the graph in later steps, which, subsequently, can also cause the
removal of vertices of layers that are not removed, making these
layers automatically smaller. Instead, we will set $\nmax = \lceil
|V| / k \rceil$, where $k$ is the length of the longest path in the graph, so
that we can hope for roughly equal numbers of vertices per layer.
We mainly used the heuristic of Coffman and
Graham~\cite{coffman1972optimal} with the minor adjustment that
preferably the leftmost layers have more vertices; we also tested
Graham's list scheduling algorithm~\cite{6767827} and an assignment with the
minimum number of layers.

\paragraph{Vertex Removal.}
After the layer assignment, the configuration usually
does not fit into the drawing area. We now
remove vertices until all vertices can be placed in the drawing area.
We first remove vertices from each layer, so that the
height of the layer is small enough. Afterwards, we remove whole
layers until the width requirement is fulfilled. We first remove
single vertices because this step can significantly influence the
total weight of layers and, therefore, the choice of layers that will
be deleted. The removal from the layers is done from left to right
since the removal of a vertex from a layer can cause other vertices
right of it to also be removed, if they become unreachable
from~$s$.

When removing from a layer, we should prefer light
vertices. However, we must also take the heights of vertices into
account: Removing a high vertex may save as much space as removing
several lighter vertices whose weight sums up to a larger value.
Hence, we measure the importance of a vertex $v$ as $i(v) =
w(v)/h_v$ and remove as many vertices of lowest importance as
necessary so that the layer fits into the drawing area. We also
tried other importance measures by taking the possible decrease
of the width of the layer or the decrease of its area into account
when removing a vertex. However, these measures did not perform
better than the simpler height-based measure.

Once all layers have a feasible height, we will remove complete
layers so that we are within the allowed total width. Note that we
must always keep gaps between adjacent layers so that the edges can be
drawn. The removal of layers is also done from left to right.
We do this based on the importance of a layer~$L$, which we define as
$i(L) = \sum_{v \in L} w(v) / \operatorname{width}(L)$, where the
width of $L$ is determined by its widest vertex.

\paragraph{Crossing Minimization.}
For the crossing minimization, we use the methods commonly used in
the Sugiyama framework, which are based on considering only (parts of)
edges between adjacent layers, but do so multiple times.
For the adjacent exchange heuristic, we also
considered the version where the weight of crossing edges is
minimized. This heuristic just performs swaps of adjacent vertices
in a layer---if this reduces the number (or weight) of
crossings. Hence, weights can easily be integrated.

\paragraph{Edge Removal.}
Even after crossing minimization, there could still be too many
crossings for the drawing to be well readable, if the graph is dense.
Hence, we add a step in
which edges are removed, if necessary. To this end, we introduce a
measure for the importance of an edge $e$: If $E'(e)$ is the set of
edges that cross $e$, then the importance of $e$ is $i(e) =
w(e)/(\sum_{e' \in E'(e)} w(e'))$. The result is that edges without a
crossing are considered most important and will never be removed.
Furthermore, an edge that crosses heavy edges---which are more
valuable to us---will more likely be removed.

\paragraph{Gaps in Layers.}
At this point, we know the orders of vertices in layers,
where a layer also includes edges that are routed through several
layers. We improve the readability by using different
gaps between the objects in a layer: Two edges will be drawn closer
together than two vertices, and edges that stay parallel when going to
the next layer can be drawn even closer. While the different sizes of
gaps make the drawings nicer, the consequence is also that only now we
know the precise height each a layer. This can, in some cases, make it
necessary to remove another vertex of a layer, which, again, is done
based on the importance measure.

\paragraph{Coordinate Assignment.}
For the final adjustment of the vertex positions, we still have some
flexibility if the vertices (and edges) in the layer do not consume
the total available height. We can use this flexibility and try to
minimize the number of edge bends.
Therefore, we integrated a part of the
heuristic of Sander~\cite{sander1996fast}. However, due to the height
constraint, we usually cannot save too many bends.

\paragraph{Edge Routing.}
Finally, only the edges need to be drawn. There are several
subproblems that we will only briefly review:
\begin{compactitem}
  \item We have to distribute the ports of the edges at the incident
    vertices or use a single port shared by all edges.
  \item We indicate the weight of edges by drawing them with
    different width. Since there are only few edges that are very
    heavy, it makes sense to not use a linear dependency between width
    and height but, e.g., a logarithmic dependency, or a dependency to
    the cube root (which gave the nicest results for our drawings).
  \item We have to distribute the vertical segments of the edges
    between consecutive layers. We want to do this such that
    both overlaps between segments and unnecessary (double) crossings
    are avoided. We first find a relative order of the
    segments from left to right for each pair of adjacent layers as
    follows: For any pair of edges between the layers that do not have
    to cross, there is at most one order with an
    unnecessary crossing. Using these orders, we build a directed
    graph of vertical segments. A topological sorting of this graph
    yields an order of the segments that avoids all unnecessary
    crossings.

    Once a relative order of the vertical segments is found, we can
    assign the final coordinates. Several small improvements are
    possible that locally optimize the spacing between the edge
    segments and we have put a lot of effort into implementing
    some of them.  The most valuable optimization was the use of a
    force-directed algorithm with repelling forces between adjacent
    vertical segments that optimizes the distances between the
    segments.
\end{compactitem}

\subsection{Extensions}

\paragraph{Preprocessing.}
Similar to our algorithm for general graphs, we use a preprocessing
step in which very light vertices and edges are removed in order to
speed up the later steps.

\paragraph{Reinsertion of Removed Vertices.}
After crossing minimization, when the order of vertices is
fixed, it is possible that we could safely reinsert some of the
removed vertices so that the available area is used in a better way.
We prefer the vertices with the highest importance as
defined before and insert them in the leftmost available layer.
Note that after reinserting vertices, we may
have to reorder some of layers for crossing minimization.

\paragraph{B\'ezier Curves.}
While we try to avoid bends of the orthogonal edges, there still can
be longer edges that have several bends, making them hard to follow.
We suggest drawing the edges as smooth curves
instead. To this end, we represent each edge segment between two
adjacent layers as a cubic B\'ezier curve. As a simple
version, this can be done by making the two bends of the orthogonal
edge the two middle control points of the curve; this yields already quite
nice results. We can further improve the drawings by
adjusting the force-directed algorithm for the vertical segments
(i.e., the control points): First, we can allow horizontal segments
to overlap since they are not actual segments any more. Second, we can
add a tendency to put the segments close to the middle between the
layers in order to avoid sharp bends. We can, however, not place all
vertical segments in the middle; doing so could result in unnecessary
crossings of the respective curves.

\paragraph{Weight Transfer.}
Suppose we delete a vertex $v$ such that the edges
$(u,v)$ and $(v,w)$ for vertices $u$ and $w$ exist. If both edges are
heavy, it is possible that many students reached $w$ from $u$ with
$v$ as an intermediate step. Hence, after the removal, an edge
$(u,w)$ becomes more valuable to us because this edge can also
partially represent the steps described above. We can
model this by creating edge~$(u,w)$---if it did not exist---and
increasing its weight by $\min\left\{ w(u,v), w(v,w) \right\}$ for the
remainder of the algorithm. Similar weight transfers make sense also
in more complicated situations.

\subsection{Experiments and Evaluation}
Also the algorithm for calculation graphs was implemented in Java. We
used real-world data generated in user studies, with graphs of 107 and
more vertices. The largest graph had 1031 vertices and 1549 edges. As
for general graphs, we mainly used the A4 paper size as the prescribed
drawing area. The tests were performed on a 3 GHz CPU with 4 GB RAM.
Our main results are as follows.
\begin{compactitem}
  \item A preprocessing that removes the lightest vertices often
		improves the output, i.e., the drawn subgraph is heavier; see
		Table~\ref{tab:schrankenwerte} in Appendix~\ref{app:calc-tables}.
  \item In our tests, there was no significant influence of the choice
    of the layering algorithm on the final weight of the drawn
    subgraph, especially when the postprocessing for vertex
    reinsertion was used.
  \item The adjacent exchange heuristic for crossing
    minimization and taking the weight of crossing edges into
    account reduces the weight of crossing edges significantly
    (factor $> 2$) and causes only few additional crossings; see
    Table~\ref{tab:krMin} in Appendix~\ref{app:calc-tables}.
  \item The runtime requirement for the largest graph with 1031
    vertices was 3 to 4 seconds, depending on the set parameters; the
    largest part of the time was spent on the force-directed adjustment
    of segments in the edge routing step and on writing the output
    file (about 2 seconds).
\end{compactitem}

\begin{figure}[tb]
  \centering
  \includegraphics[scale=0.32, page=2]{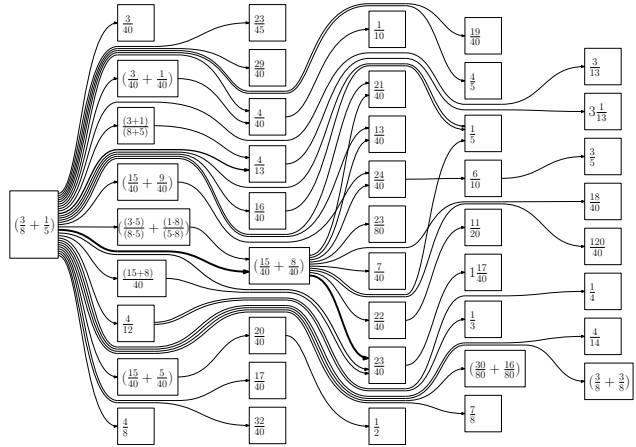}
  \caption{An output example with edges drawn as B\'ezier curves
  (scaled down).}
  \label{fig:calc-output-bezier-scaled}
\end{figure}

Figure~\ref{fig:calc-output-bezier-scaled} shows an output example
using B\'ezier curves, which, in our opinion, is the nicer and more
interesting style compared to the version with orthogonal edges. 
Full examples can be found in Appendix~\ref{app:calc-examples}.

\section{Conclusion}

We have introduced the problem of drawing a heavy subgraph in
a prescribed area. Both for general graphs without further constraints
and for the special application of calculation graphs, we have
developed and tested heuristics which yield quite nice results.



\paragraph{Acknowledgement.} We would like to thank Martin Hennecke
for introducing the problem of drawing calculation graphs to us and
for providing input data.

\bibliographystyle{splncs03}
\bibliography{abbrv,rad,cgraph}

\newpage
\appendix
\section{General Graphs}

\subsection{Pseudocode and Tables}
\label{app:gen-tables}

\begin{algorithm}[h]
  \SetKw{KWor}{or}
  compute equilibrium layout\;
  initialize frame~$F$ as bounding box around the drawing\;
	\While{frame~$F$ too large}{
    shrink $F$ by constant value\;
    push vertices that are (partially) outside of the frame
    into~$F$\;
    compute new equilibrium layout respecting the boundary
    of~$F$\;
    \While{min.\ edge length $< \ladj$ \KWor min.\ distance of
      nonadjacent vertices $< \lnadj$}{
        \eIf{average edge length $\le \lunit \cdot \clen$}{
        remove vertex $v$ with maximum stress $s(v)$}{
        remove edge $e$ with maximum stress $s(e)$}
      }
    compute new equilibrium layout respecting the boundary
    of~$F$\;
	}
	\caption{Algorithm for general graphs}
	\label{alg:general-graphs}
\end{algorithm}

\begin{table}[h]
  \centering
    \begin{tabular}{|c|r|r|r|r|r|r|}\hline
    \multirow{2}*{\cpre} & \multicolumn{2}{c|}{vertices} &
    \multicolumn{2}{c|}{edges} & \multicolumn{2}{c|}{runtime
    (s)}\\\cline{2-7}
    & weight & change & weight & change & runtime & change\\\hline
	0.8 & 659.4 & --- & 716.4 & --- & 111.7 & ---\\
	0.75 & 670.4 & + 1.7 \% & 756.4 & + 5.6 \% & 135.8 & + 21.6 \%\\
	0.7 & 675.4 & + 2.4 \% & 769.6 & + 7.4 \% & 163.2 & + 46.1 \%\\	
	0.65 & 668.6 & + 1.4 \% & 768.0 & + 7.2 \% & 242.1 & + 116.7 \%\\	
	0.6 & 667.8 & + 1.3 \% & 745.2 & + 4.0 \% & 708.7 & + 534.5 \%\\\hline		
  \end{tabular}

  \smallskip

  \caption{Effect of the factor \cpre on runtime and quality of the
    results for the complete publication graph.}
  \label{tab:cpre}
\end{table}

\begin{table}[h]
  \centering
    \begin{tabular}{|c|r|r|r|r|r|r|}\hline
    \multirow{2}*{$f_{\text{precision}}$} &
    \multicolumn{2}{c|}{vertices} & \multicolumn{2}{c|}{edges} &
    \multicolumn{2}{c|}{runtime (s)}\\\cline{2-7}
    & weight & change & weight & change & runtime & change\\\hline
    0.7 & 687.7 & 0 & 719.0 & 0 & 144.4 & 0\\
	0.8 & 664.1 & -- 3.4 \% & 741.2 & + 3.1 \% & 149.0 & + 3.2 \%\\
	0.9 & 673.5 & -- 2.1 \% & 742.6 & + 3.3 \% & 127.6 & -- 11.6 \%\\
	1.0 & 660.7 & -- 3.9 \% & 739.6 & + 2.9 \% & 139.1 & -- 3.7 \%\\
	1.1 & 628.5 & -- 8.6 \% & 705.0 & -- 1.9 \% & 133.7 & -- 7.4 \%\\\hline
  \end{tabular}

  \smallskip

  \caption{Effect of the choice of the parameter $\clen$ on runtime
    and quality of the results for the complete publication graph.}
  \label{tab:clen}
\end{table}

\begin{table}[h]
  \centering
    \begin{tabular}{|c|r|r|r|r|r|r|}\hline
    \multirow{2}*{} & \multicolumn{2}{c|}{vertices} &
    \multicolumn{2}{c|}{edges} & \multicolumn{2}{c|}{runtime (s)}\\\cline{2-7}
    & weight & change & weight & change & runtime & change\\\hline
	1 x equilibrium & 663.1 & 0 & 412.4 & 0 & 107.1 & 0\\
	2 x equilibrium & 666.7 & + 0.5 \% & 739.6 & + 79.3 \% & 124.9 & + 16.6 \%\\\hline
  \end{tabular}

  \smallskip

  \caption{Improvement in terms of the total edge weight by computing
  the equilibrium layout always twice, first without and then with the
  repelling force between vertices and edges.}
  \label{tab:doubleCalculation}
\end{table}

\newpage

\subsection{Output Examples}
\label{app:gen-output}

In our examples, vertex weights are indicated using colors from red
(heavy) to green (light).

\begin{sidewaysfigure}[htbp]
	\centering
	\includegraphics[width=0.9\textheight]{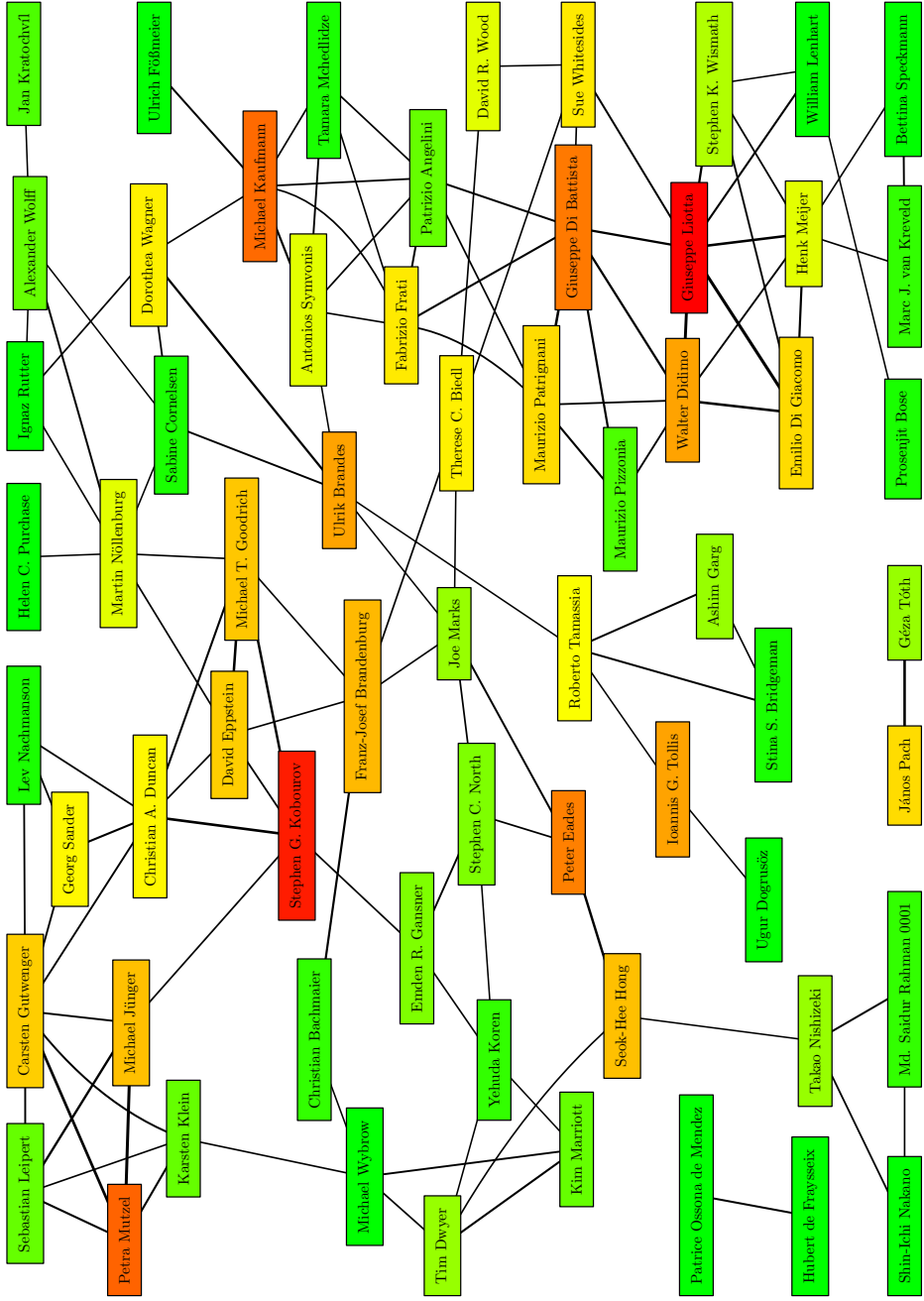}
		\caption{Example for the complete publication graph with
      $\lunit=3.0\text{ cm}$; vertices contained in drawing: 6.6 \%;
      vertex weight contained in drawing: 42.1 \%; edges contained in
      drawing: 4.9 \%; edge weight contained in drawing: 17.0 \% }
	\label{fig:gen-result2}
\end{sidewaysfigure}

\begin{sidewaysfigure}[htbp]
	\centering
	\includegraphics[width=0.9\textheight]{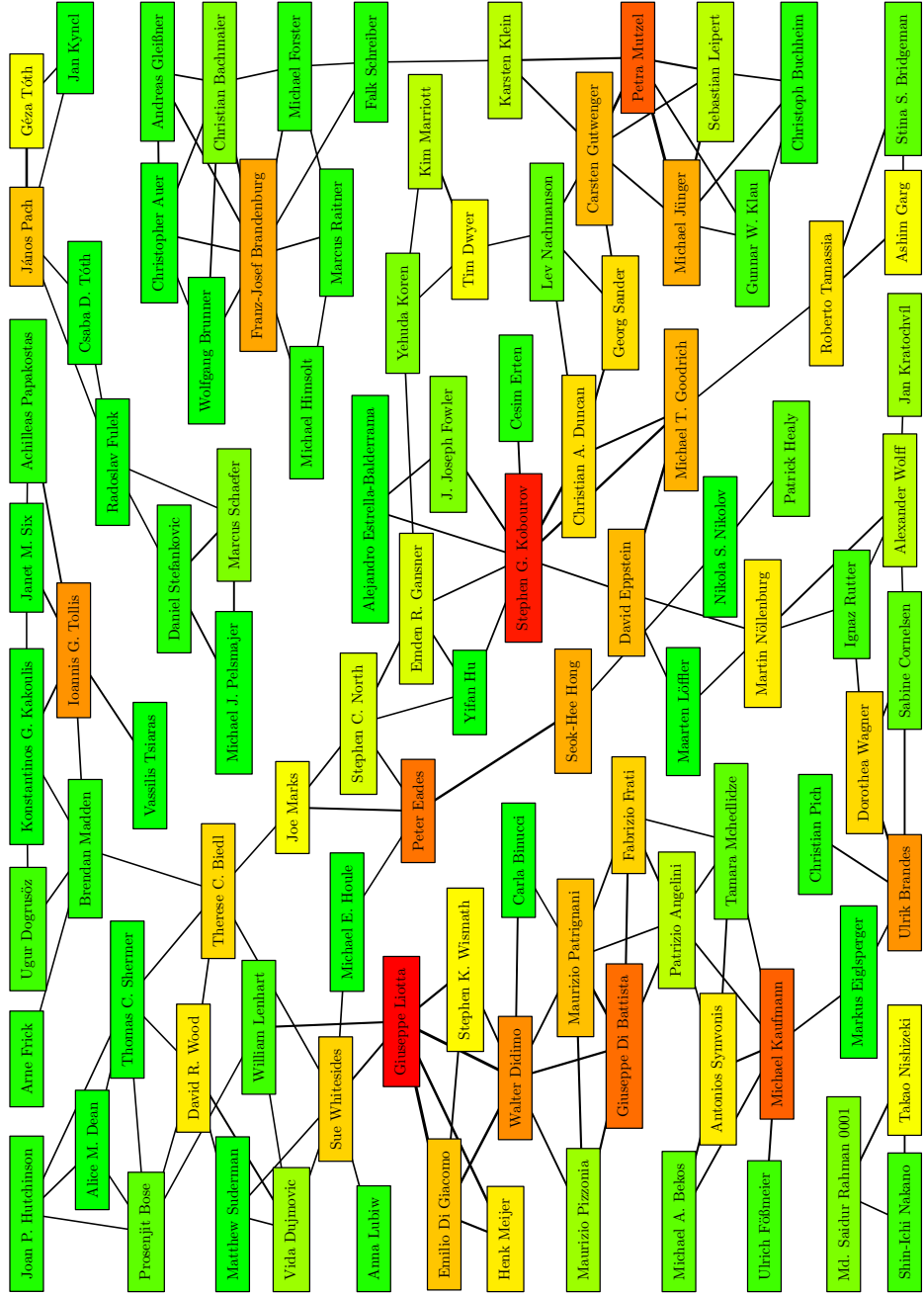}
		\caption{Example for the complete publication graph with
      $\lunit=2.0\text{ cm}$; vertices contained in drawing: 10.1 \%;
      vertex weight contained in drawing: 49.4 \%; edges contained in
      drawing: 6.9 \%; edge weight contained in drawing: 21.2 \%}
	\label{fig:gen-result3}
\end{sidewaysfigure}

\begin{sidewaysfigure}[htbp]
	\centering
	\includegraphics[width=0.9\textheight]{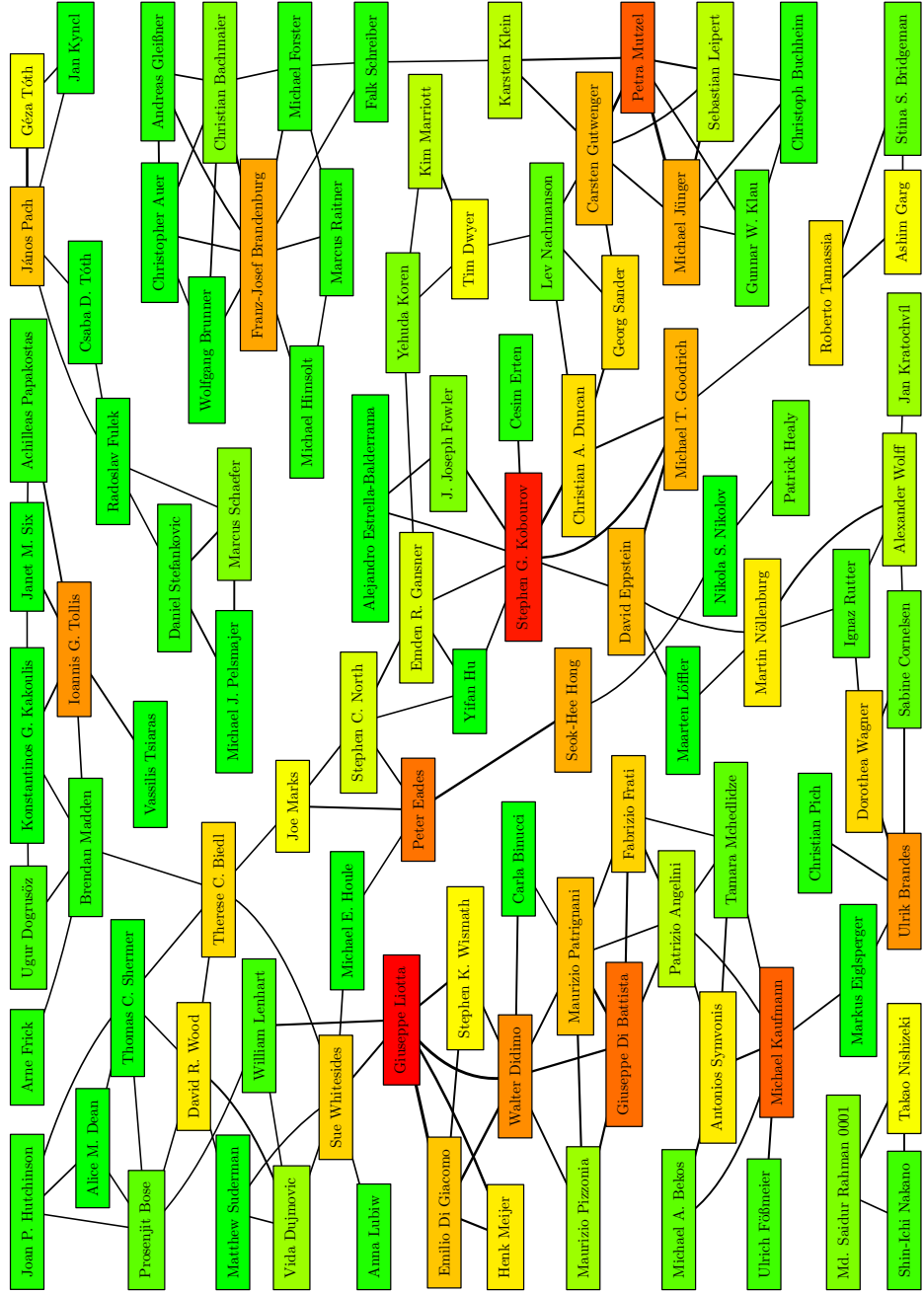}
	\caption{Same drawing as shown in Fig.~\ref{fig:gen-result2}, but
	some edges are transformed into quadratic B\'ezier curves.}
	\label{fig:gen-result1}
\end{sidewaysfigure}

\begin{sidewaysfigure}[htbp]
	\centering
	\includegraphics[width=0.9\textheight]{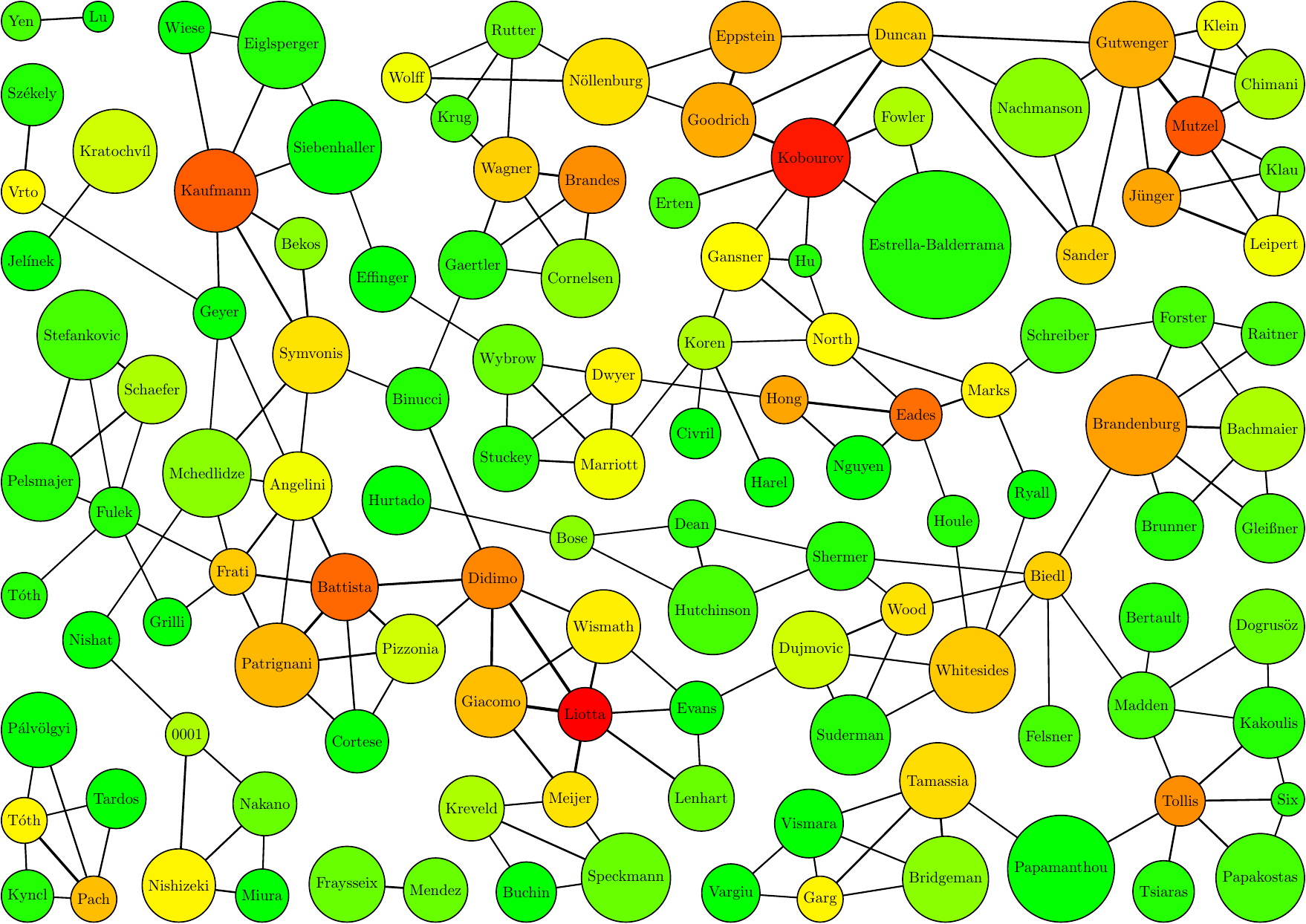}
		\caption{Example for the complete publication graph (only last
      part of the name) with $\lunit=2.0\text{ cm}$; vertices contained
      in drawing: 12.8 \%; vertex weight contained in drawing: 54.2
      \%; edges contained in drawing: 8.6 \%; edge weight contained in
      drawing: 24.0 \%}
	\label{fig:gen-result5}
\end{sidewaysfigure}
\newpage

\section{Calculation Graphs}
\label{app:calc}

\subsection{Tables}
\label{app:calc-tables}

\begin{table}
\centering
    \begin{tabular}{|r|r|r|r|r|r|r|}
		\hline
    ~ & \multicolumn{2}{r|}{vertex weight} &
		\multicolumn{2}{r|}{edge weight} \\
    \hline
		number of vertices & threshold & improvement & threshold &
		improvement \\
    \hline
		107 & 7 & 4.49 \% & 8 & 7.15 \% \\
    143 & 2 & 0.45 \% & 2 & 0.57 \% \\
    170 & 7 & 9.21 \% & 8 & 14.80 \% \\
		203 & 2 & 2.24 \% & 2 & 3.20 \% \\
		331 & 4 & 3.32 \% & 4 & 6.17 \% \\
		358 & 4 & 1.92 \% & 4 & 3.12 \% \\
		406 & 14 & 7.12 \% & 14 & 11.33 \% \\
		530 & 7 & 6.49 \% & 7 & 11.26 \% \\
    563 & 7 & 9.32 \% & 8 & 16.60 \% \\
    663 & 8 & 4.96 \% & 8 & 9.33 \% \\
		755 & 9 & 6.76 \% & 7 & 9.50 \% \\
		775 & 8 & 5.02 \% & 8 & 7.87 \% \\
		799 & 10 & 10.35 \% & 10 & 17.28 \% \\
		1031 & 11 & 6.79 \% & 10 & 8.64 \% \\
		\hline
    \end{tabular}

  \smallskip

\caption{Improvement in the vertex and edge weight when activating the
preprocessing that removes vertices and edges whose weight is below
the given threshold. The table shows results for input graphs of
different size.}
\label{tab:schrankenwerte}
\end{table}

\begin{table}
  \centering
    \begin{tabular}{|c|r|r|r|r|}
    \hline
    ~& vertex weight & edge weight & crossings & weight of crossings \\
    \hline
    adjacent exchange heuristic & 6841.25 & 6447.48  & 14.41 & 3041.54 \\
    adjacent exchange with weights & 6843.14 & 6459.70 & 15.10 & 1219.68 \\
    median heuristic & 6845.78 & 6463.01 & 13.65 & 3114.88 \\
    \hline
    \end{tabular}

  \smallskip

	\caption{Comparison of the heuristics for crossing minimization.
	Resulting values are averages.}
  \label{tab:krMin}
\end{table}

\subsection{Output Examples}
\label{app:calc-examples}

\begin{sidewaysfigure}[htbp]
\centering
  \includegraphics[scale=0.75]{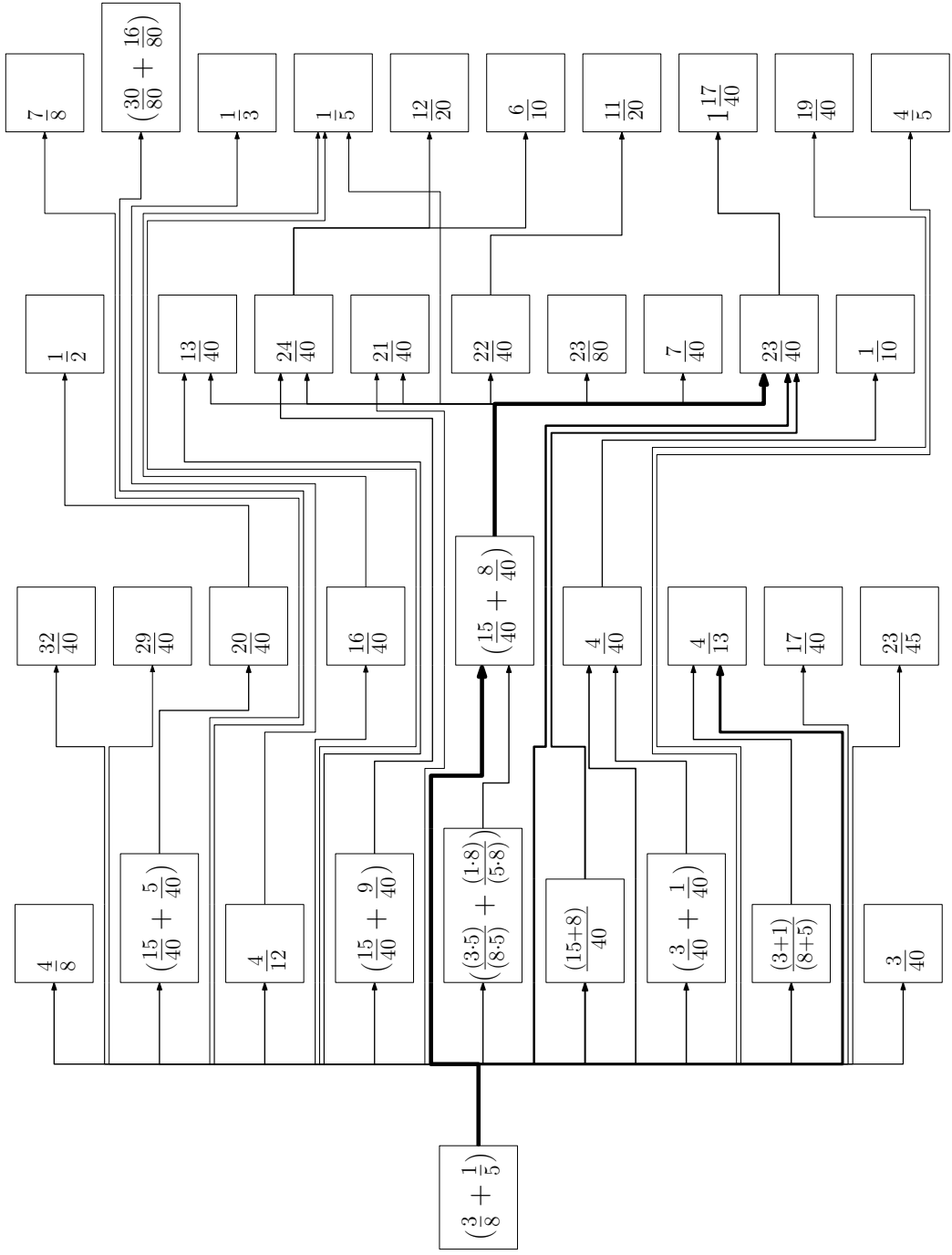}
  \caption{Orthogonal edges with a single port per vertex.}
	\label{fig:calc-example-single-port}
\end{sidewaysfigure}

\begin{sidewaysfigure}[htbp]
\centering
  \includegraphics[scale=0.75, page=1]{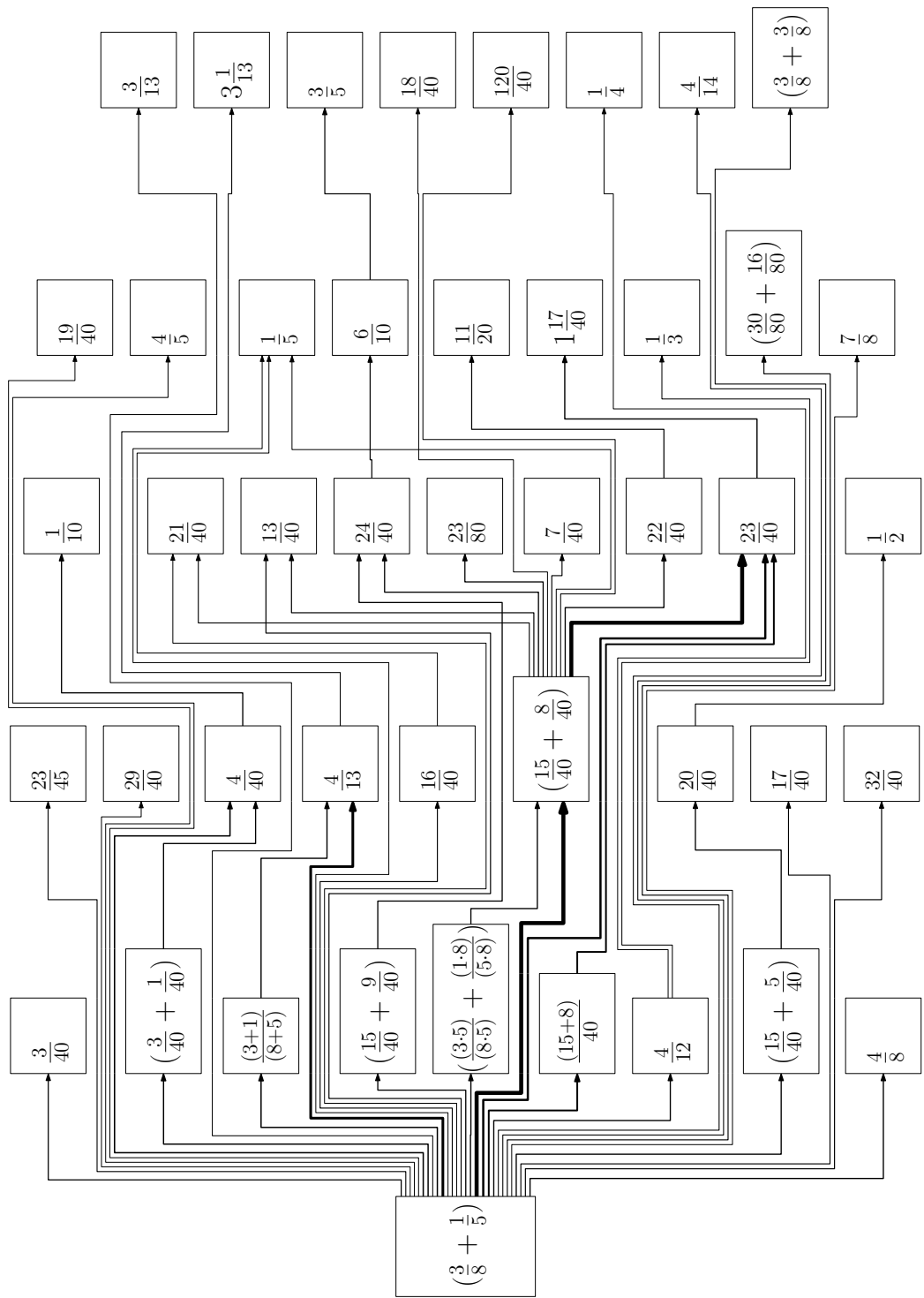}
  \caption{Orthogonal edges with an own port for each outgoing edge.}
	\label{fig:calc-example-multiple-ports}
\end{sidewaysfigure}

Here, we have included some output drawings of calculation graphs,
showing different drawing styles for the edges. In
Fig.~\ref{fig:calc-example-single-port}, we have orthogonal edges, but
only a single port per vertex shared by all outgoing edges. In
Fig.~\ref{fig:calc-example-multiple-ports}, the edges are also drawn
orthogonally, but now each edge has its own port. Finally,
Fig.~\ref{fig:calc-example-multiple-ports} shows edges that are drawn
as B\'ezier curves. The input graph contained 358~vertices, of which
38 are still contained in Fig.~\ref{fig:calc-example-single-port},
i.e., only 10.6 \%. However, these vertices represent 91.1 \% of the
total vertex weight.

\begin{sidewaysfigure}[htbp]
\centering
  \includegraphics[scale=0.75, page=2]{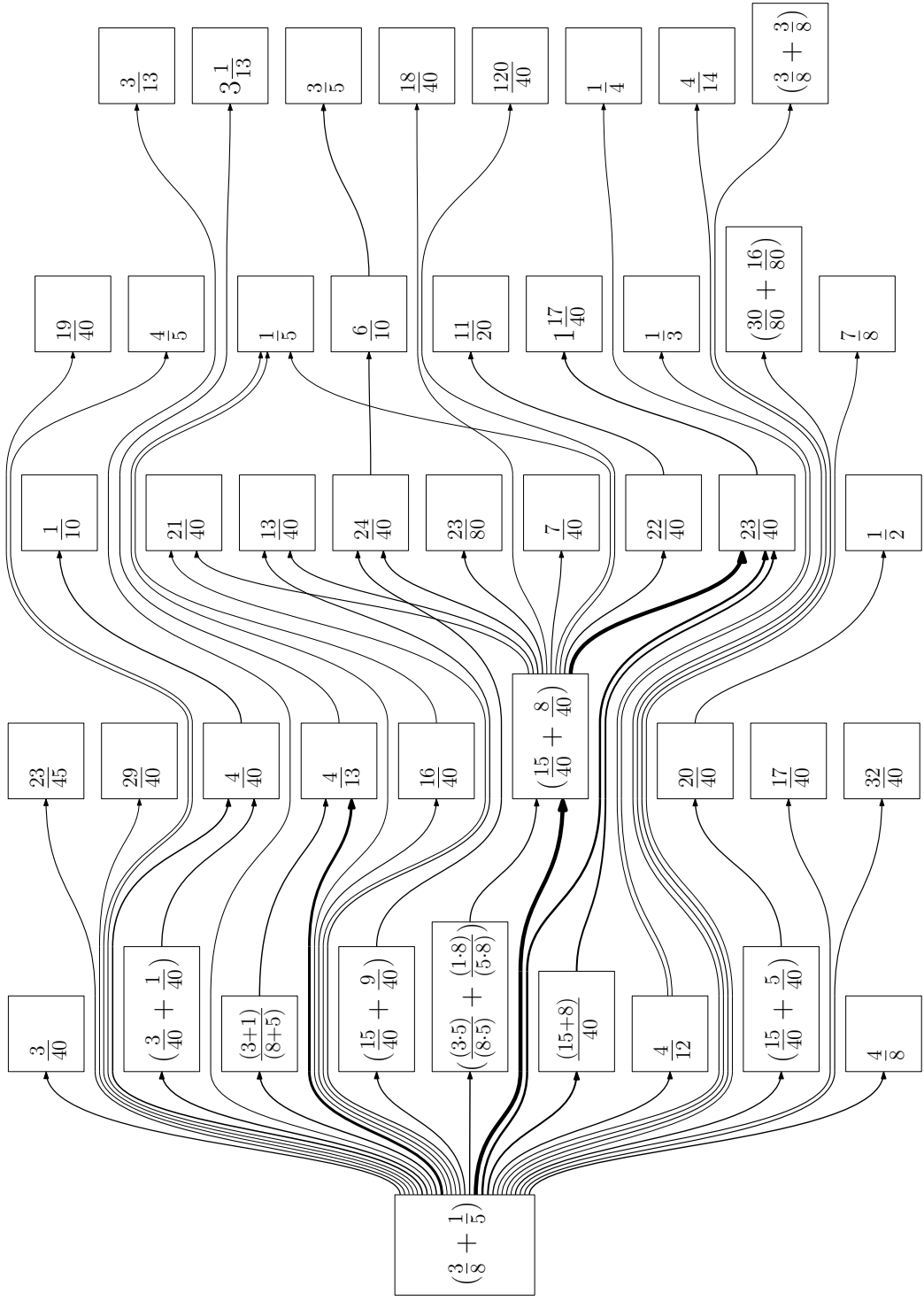}
  \caption{Edges drawing as B\'ezier curves; except for the edge
    style, the drawing is the same as in
    Fig.~\ref{fig:calc-example-multiple-ports}.}
	\label{fig:calc-example-bezier}
\end{sidewaysfigure}

\begin{sidewaysfigure}[htbp]
\centering
  \includegraphics[scale=0.6]{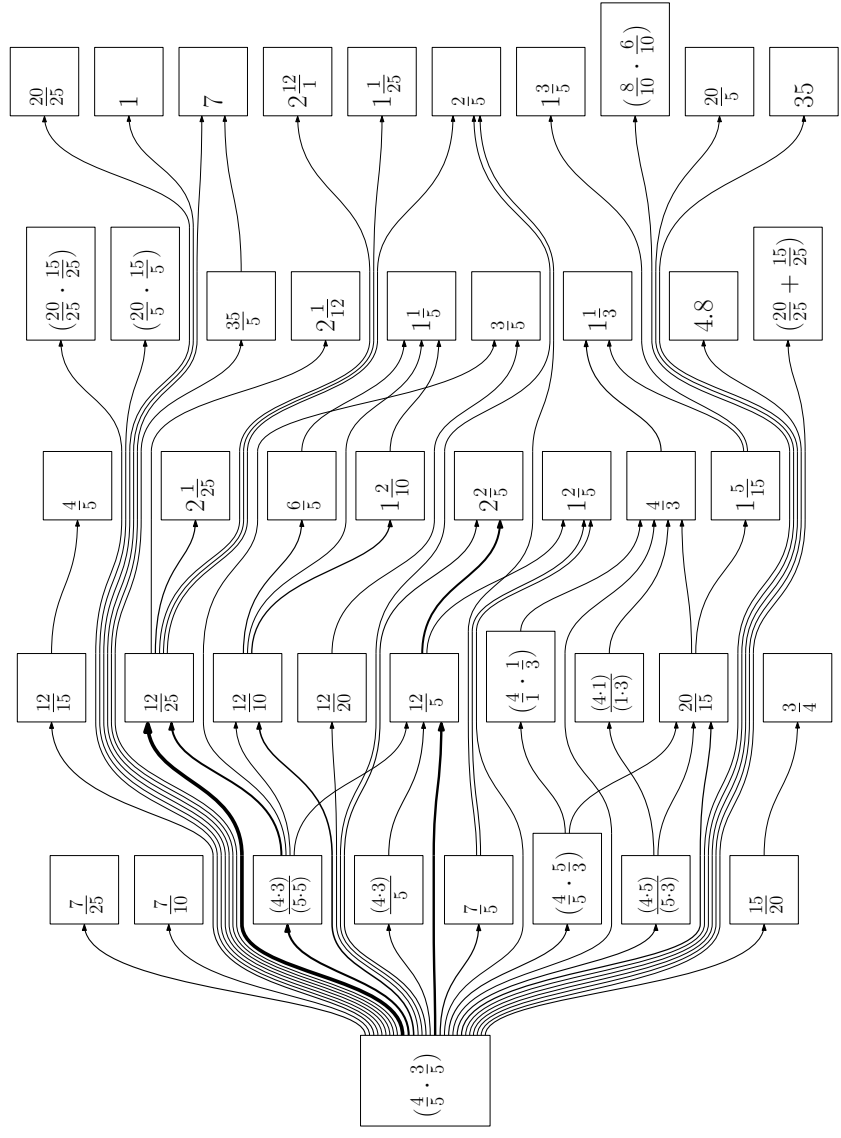}
  \caption{Output drawing for a graph with 203 vertices; contained
	vertices: 22.1 \%; contained vertex weight: 95,2 \%; contained
edges: 22.0 \%; contained edge weight: 94,0 \%}
\end{sidewaysfigure}

\begin{sidewaysfigure}[htbp]
\centering
  \includegraphics[scale=0.75]{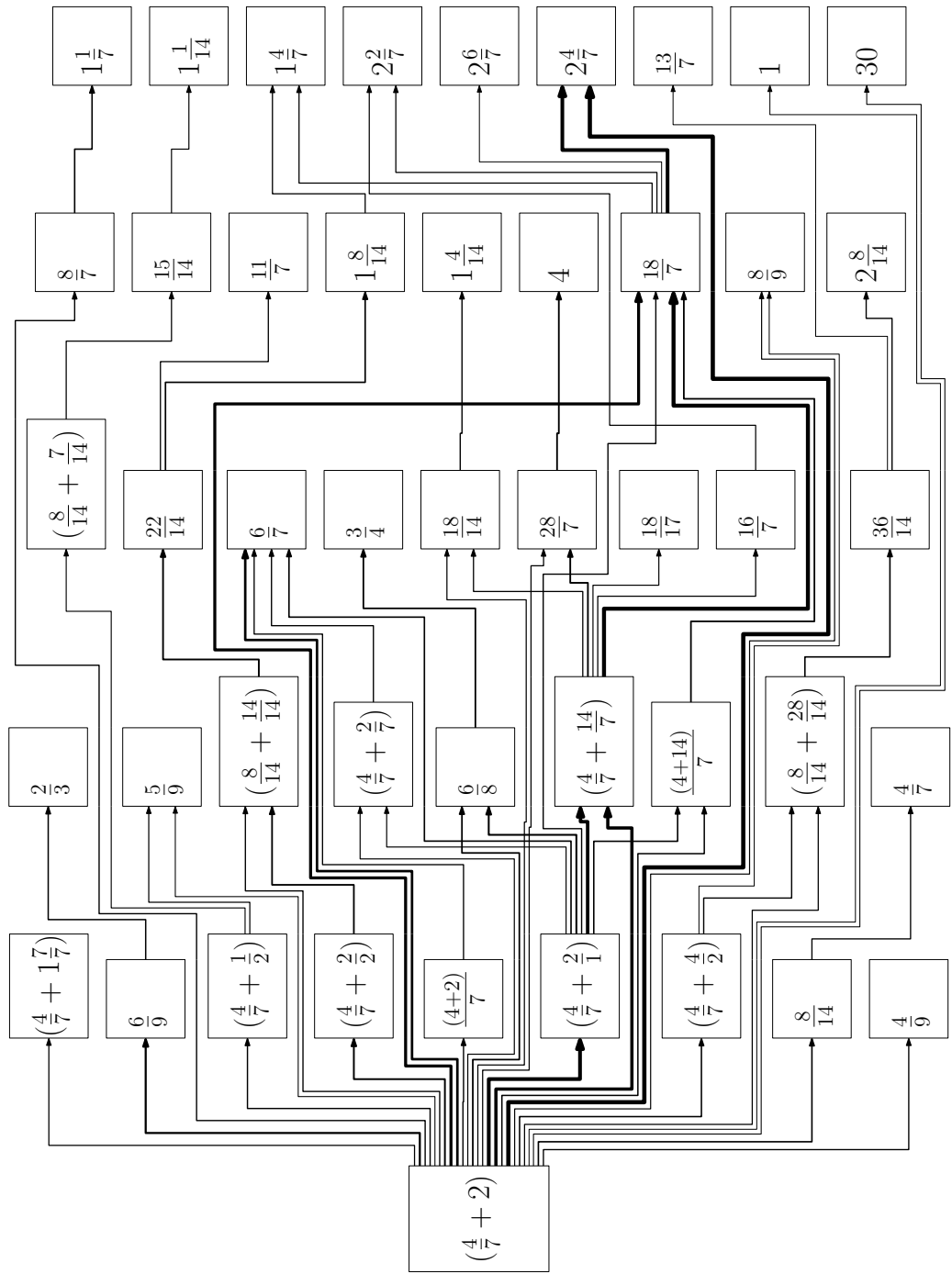}
  \caption{Output drawing for a graph with 331 vertices; contained
	vertices: 13,9 \%; contained vertex weight: 92,9 \%; contained
edges: 11,4 \%; contained edge weight: 89,6 \%}
\end{sidewaysfigure}

\begin{sidewaysfigure}[htbp]
\centering
  \includegraphics[scale=0.75]{ausgabe775multi}
  \caption{Output drawing for a graph with 775 vertices; contained
	vertices: 5,8 \%; contained vertex weight: 80,6 \%; contained edges:
4,6 \%; contained edge weight: 75,0 \%}
\end{sidewaysfigure}

\begin{sidewaysfigure}[htbp]
\centering
  \includegraphics[scale=0.75]{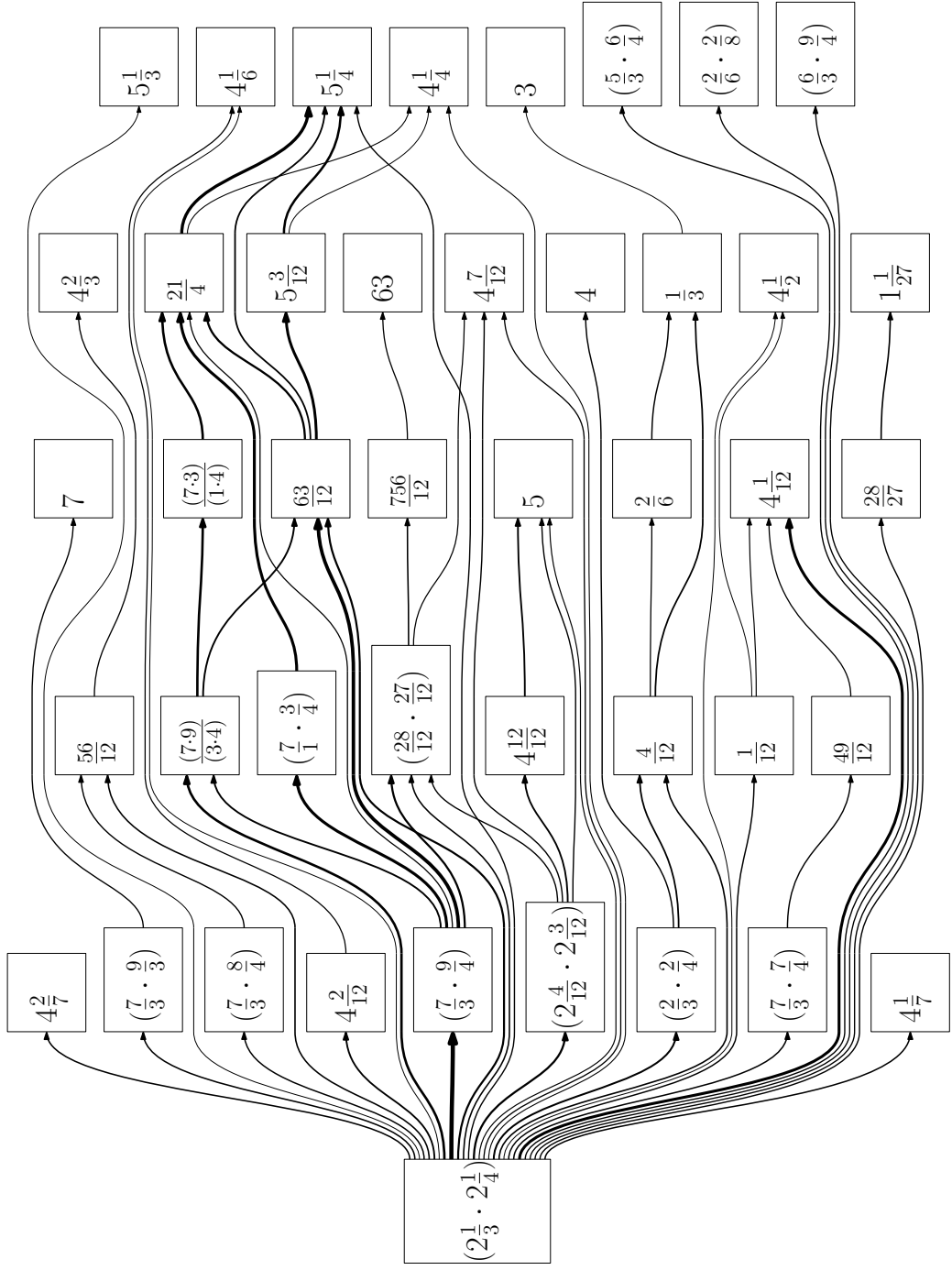}
  \caption{Output drawing for a graph with 799 vertices; contained
	vertices: 5,4 \%; contained vertex weight: 81,7 \%; contained edges:
5,3 \%; contained edge weight: 78,7 \%}
\end{sidewaysfigure}

\begin{sidewaysfigure}[htbp]
\centering
  \includegraphics[scale=0.75]{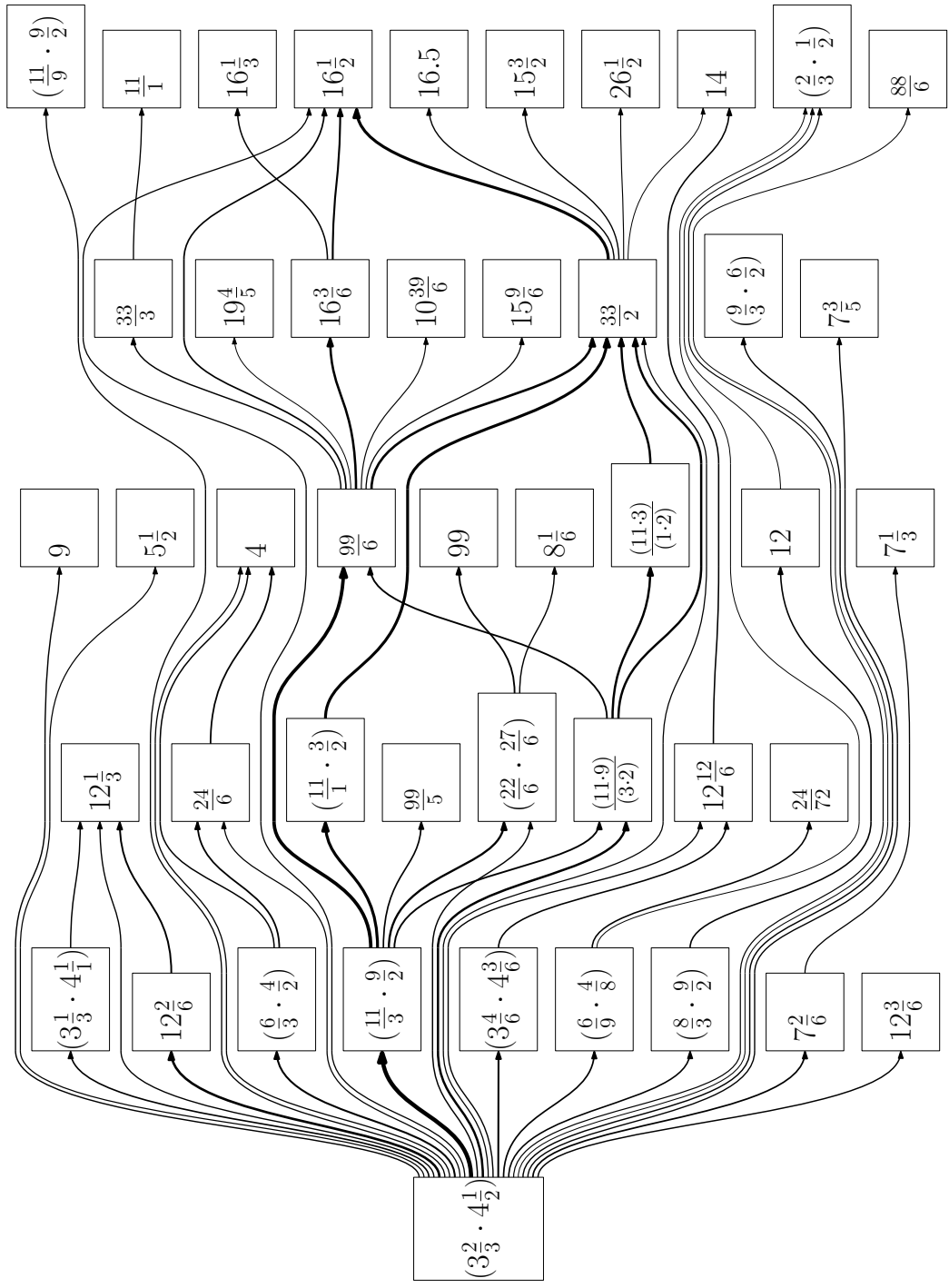}
  \caption{Output drawing for a graph with 1031 vertices; contained
	vertices: 4,7 \%; contained vertex weight: 75,9 \%; contained edges:
4,1 \%; contained edge weight: 74,1 \%}
\end{sidewaysfigure}

\begin{sidewaysfigure}[htbp]
\centering
  \includegraphics[scale=0.75]{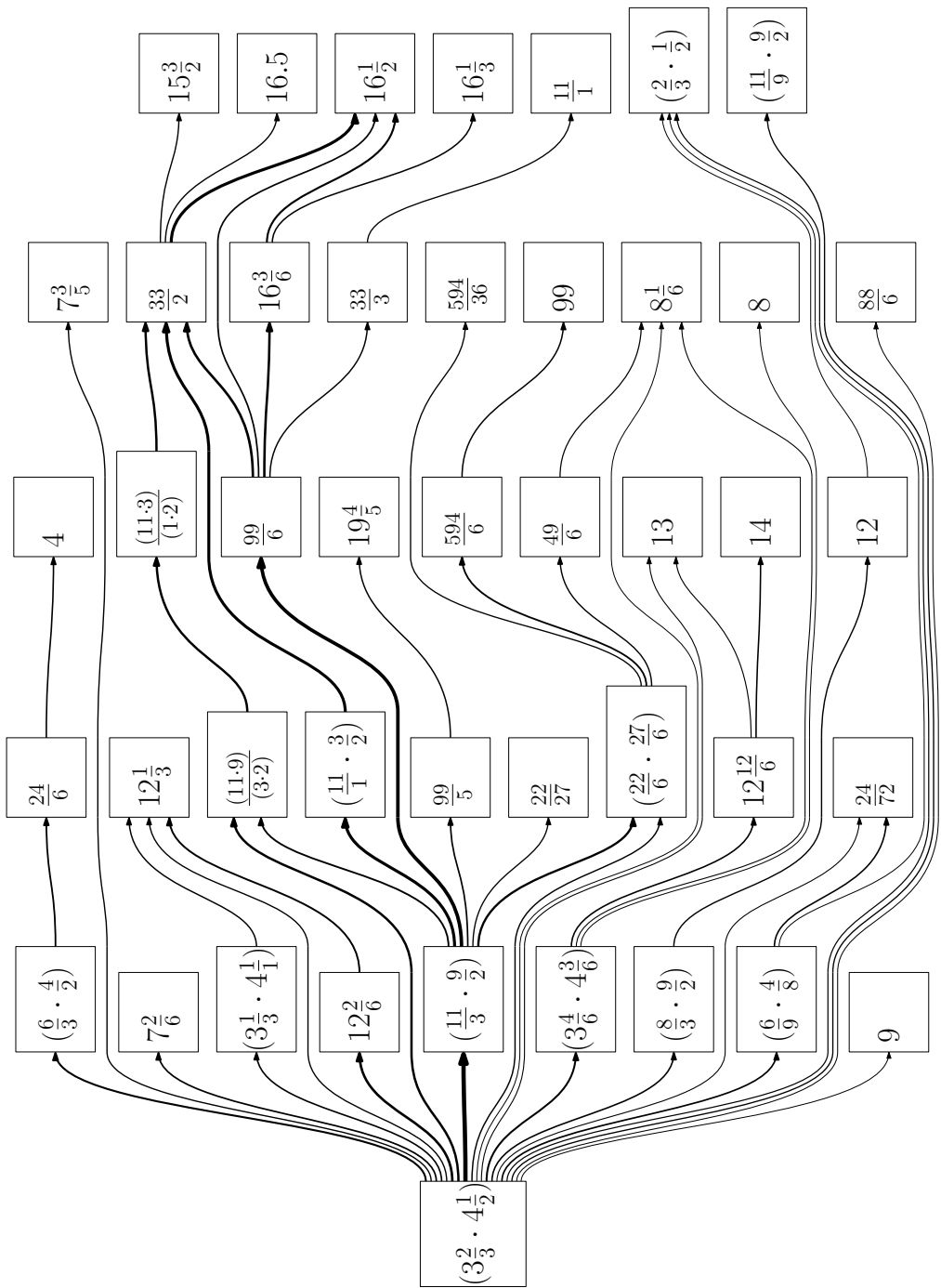}
  \caption{Output drawing for a graph with 1031 vertices; no crossings
		are allowed; contained vertices: 4,3 \%; contained vertex
weight: 76,6 \%; contained edges: 3,9 \%; contained edge weight: 70,2 \%}
\end{sidewaysfigure}

\end{document}